\documentclass{PoS}

\PoS{PoS(HEP2005)320}

\title{A Search for Charged Massive Stable Particles}

\ShortTitle{A Search for Charged Massive Stable Particles}

\author{\speaker{Thomas Nunnemann}\thanks{for the D\O{} Collaboration}\\
        LMU Munich, Germany\\
        E-mail: \email{Thomas.Nunnemann@physik.uni-muenchen.de}}

\abstract{A search for charged massive (quasi-) stable particles with the D\O{}
detector at the Tevatron collider based on 390\,pb$^{-1}$
of data is presented. 
The search is performed in the frameworks of gauge-mediated supersymmetry
breaking and the minimal supersymmetric extentension of the standard model.
The hypothetical particles are
assumed to be pair-produced in $p\bar{p}$ collisions giving a signature
of two reconstructed muon-like objects with high invariant mass and 
time-of-flights indicative of heavy particles. 
Since no excess over background
is observed, cross-section limits for the pair-production of stable staus 
and
charginos are set. Mass limits of 140\,GeV for a higgsino-like chargino and
174\,GeV for a gaugino-like chargino are set.}

\FullConference{International Europhysics Conference on High Energy Physics\\
		 July 21st - 27th 2005\\
		 Lisboa, Portugal}

\begin{document}

\section{Introduction}
We report on the search for new charged massive stable particles (CMSP)
assumed to be pair-produced in $p\bar{p}$ collisions. Herein, stable refers to
a lifetime long enough to escape the entire detector without decaying.
Heavy particles can be identified with the D\O{} detector primarily using the 
time-of-flight measurement of the muon detector's scintillators, which have
a time resolution of 2-3\,ns.

Several supersymmetric (SUSY) models can include a long-lived, quasi-stable
particle
as next-to-lightest supersymmetric particle (NLSP) provided that it is nearly
mass-degenerate with the lightest supersymmetric particle (LSP) or that its
coupling to the LSP is small.

The latter condition can be fulfilled in models with gauge-mediated 
supersymmetry breaking (GMSB), where gauge interactions with messenger fields
at a scale much smaller than the Planck scale are responsible for the SUSY
breaking~\cite{Giudice}. 
GMSB models have a very distinctive phenomenology. The gravitino $\tilde{G}$
is typically light
($\lesssim 1\,\mathrm{keV}$)
and is the lightest SUSY particle
(LSP). The next-to-lightest SUSY particle (NLSP) is usually either the lightest
neutralino $\tilde{\chi}_{1}^{0}$, decaying into $\gamma\tilde{G}$,
or the lightest charged slepton (mostly $\tilde{\tau}_1$), decaying into
$l\tilde{G}$. Due to the weak gravitational coupling the NLSP can be 
quasi-stable.
In this analysis, the GMSB model referred to as ``Snowmass Model Line D'' 
which contains a stau
as NLSP is used~\cite{Snowmass}. 
Its parameter set is shown in Table~\ref{table1}.

In anomaly-mediated supersymmetry breaking models the lightest chargino
$\tilde{\chi}_1^\pm$ (NLSP) is nearly mass-degenerate with the lightest
neutralino $\tilde{\chi}_1^0$ (LSP)~\cite{Randall}. 
Two general cases have been studied in this
analysis in the framework of the MSSM (see Table~\ref{table2})
\cite{Martin}: higgsino-like
$\tilde{\chi}_1^\pm$, $\tilde{\chi}_1^0$ ({\em deep-higgsino region}) and
gaugino-like $\tilde{\chi}_1^\pm$, $\tilde{\chi}_1^0$ (AMSB inspired).

For both the GMSB and chargino analysis, signal events were generated with
Pythia 6.202~\cite{Pythia} 
and passed through a parameterized Monte Carlo simulation, which
includes all efficiencies and detector resolution smearing. In particular
the muon system's time measurements are smeared according to resolutions and
offsets measured in data. Only the pair-production of the lightest 
$\tilde{\tau}$ or the chargino, respectively, is considered. However, the
analysis has also some sensitivity to CMSPs produced in cascade-decays of
heavier supersymmetric particles.

The data sample consists of 390\,pb$^{-1}$.
The trigger employed required two tracks in the muon system using an
asymmetric trigger gate to be efficient for particles travelling slower 
than light.
Nevertheless the trigger efficiency drops to about 75\% for 
$M_{\tilde{\tau}}=300$\,GeV.

\begin{table}[h]
\begin{tabular}{lcccccc}
\hline\hline
Model & $\Lambda_m$ [GeV] & $M_m$ & $N_5$ & $\tan\beta$ & sgn $\mu$ & 
$C_{grav}$\\
\hline
GMSB ``Model Line D'' & from 19 to 100\,TeV & $2\Lambda_m$ & 3 & 15 & +1 & 1\\
\hline\hline
\end{tabular}
\caption{GMSB ``Model Line D'' parameter}
\label{table1}
\end{table}

\begin{table}[h]
\begin{tabular}{lcccccc}
\hline\hline
Model & $\mu$ [GeV] & $M_1$ [GeV] & $M_2$ [GeV] & $M_3$ [GeV] 
& $\tan\beta$ & $M_{\tilde{q}}$ [GeV]\\
\hline
higgsino-like &from 60 to 300&100,000&100,000&500&15&800\\
gaugino-like &10,000 & $3 M_2$& from 60 to 300&500&15&800\\
\hline\hline
\end{tabular}
\caption{MSSM parameter sets for chargino analysis}
\label{table2}
\end{table}

\section{Selection and Background Estimation}
The analysis requires two reconstructed muons with transverse momentum 
$p_T>15$\,GeV, both matched to a central track and with hits in at least two 
of the three scintillator layers. Cosmic veto cuts are applied and at least
one muon is required to be isolated to suppress background from 
heavy-flavor production. To ensure a good speed measurement for each particle
in the event, consistent time information in all scintillator layers are
required.
No separation with a $\mathrm{d}E/\mathrm{d}x$ measurement in the central
fiber tracker is attempted due to its small number of layers and small
photon statistics from scintillation. After this preselection the background
is largely dominated by $Z$ and Drell-Yan production.

\begin{figure}
\begin{center}
\mbox{\includegraphics[width=.5\textwidth]{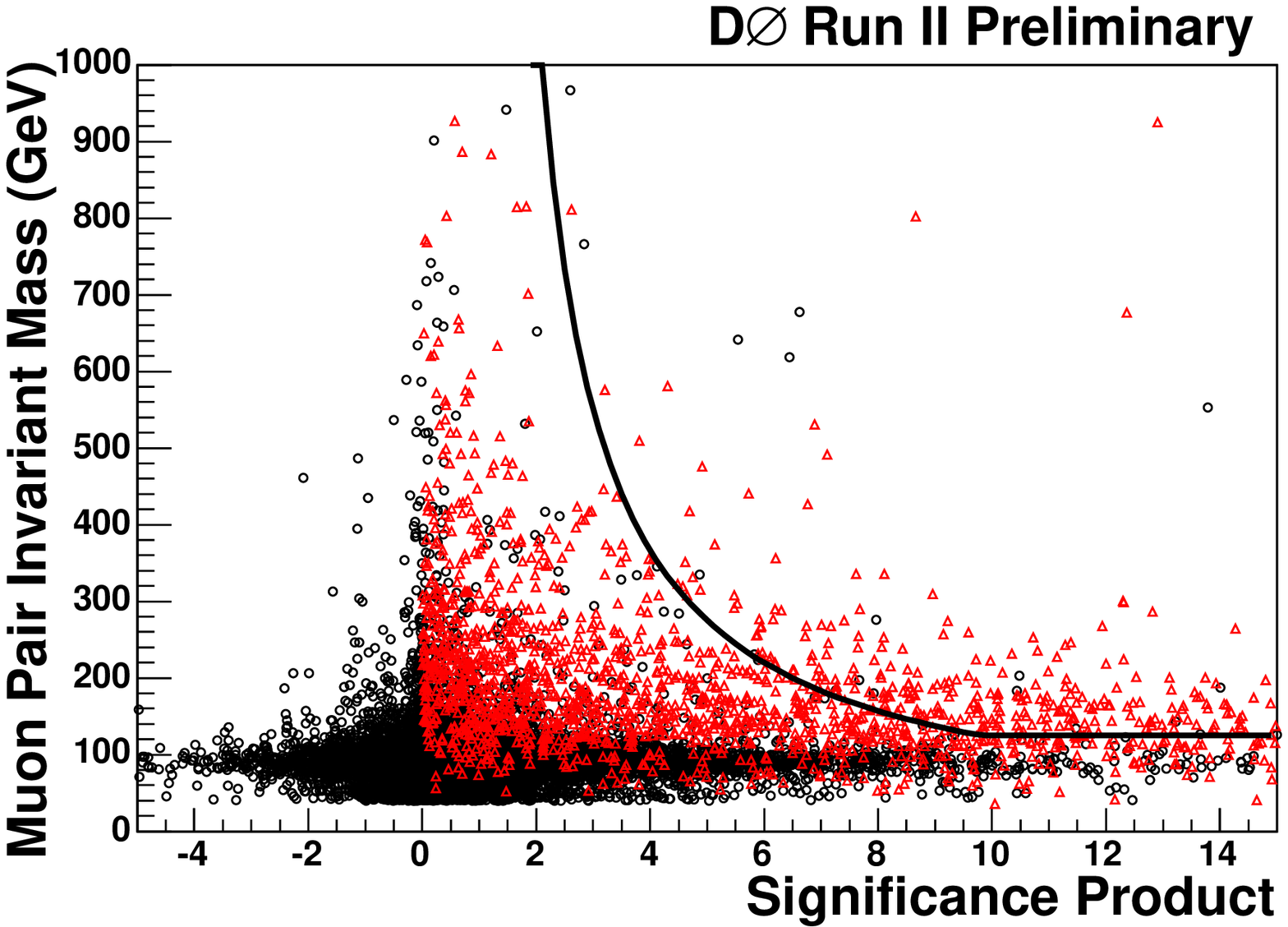}
\includegraphics[width=.5\textwidth]{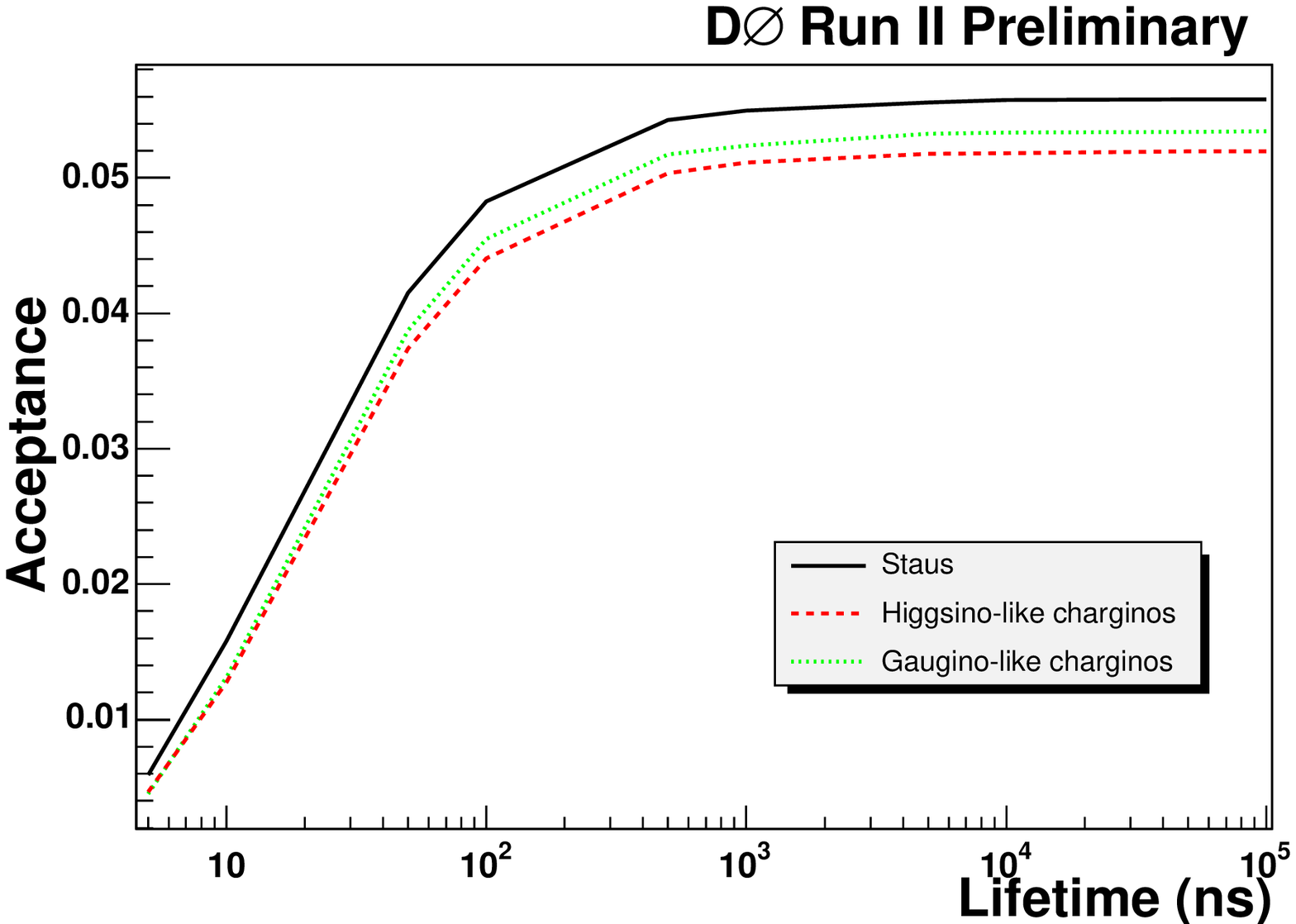}}
\end{center}
\caption{{\it Left:} Invariant mass versus significance product for reconstructed
muon pairs in data (black circles) and stau pairs with 
$M_{\tilde{\tau}}=60\,\mathrm{GeV}$ (red triangles).
The line represents the optimized two-dimensional cut. 
{\it Right:} Acceptance versus lifetime for assumed $M_{\tilde{\tau}}=100\,GeV$.}
\label{fig1}
\end{figure}

To quantify the deviation of a particle's speed $v$ from the speed of light 
$c$, a speed significance is defined as $(c-v)/\sigma_v$ which is required to
be positive.
The final signal selection is based on a two-dimensional cut using the
invariant di-muon mass $M_{\mu\mu}$ and the product of the speed significance
of the two
reconstructed muons (CMSP candidates). This cut is optimized separately
for each CMSP mass hypothesis (see Fig.~\ref{fig1}, left). Since the kinematic
properties of the considered signal models are similar, the optimization
obtained with the GMSB model is used for the other models as well.

The background estimation is entirely based on data using orthogonal data
sets to describe the two-dimensional probability density function (PDF)
depending on the
invariant di-muon mass and the significance product.
Since no correlation between these two variables is observed for the 
background, which is predominantly $Z$ and Drell-Yan production, the PDF
can be factorized. Events with apparent muon speeds $v_{\mu_{1,2}}>c$
are used to estimate the background's invariant di-muon mass distribution
and events
with $M_{\mu\mu}\approx M_Z$ are taken to estimate the significance product.

The systematic error on the signal acceptance is dominated by uncertainties in
the trigger, muon identification and timing cut efficiencies as implemented
in the simulation, and amounts to 2.7\%.
The systematic error on the background prediction has been evaluated to be
3.7\% by varying the selection criteria.

\section{Results and Conclusions}
For both the GMSB and the chargino analysis, six CMSP mass points from 60 to 
300\,GeV have been studied. The data are consistent with the background
estimation. For assumed CMSP masses larger than 100\,GeV, no events are
observed, with background predictions ranging from 0.5 to 0.7 events.
Signal acceptances are between 2.3\% and 12.3\% depending on CMSP mass and 
model.
For finite CMSP life-times the acceptance decreases as shown in 
Fig.~\ref{fig1}, right.

\begin{figure}
\includegraphics[width=.5\textwidth]{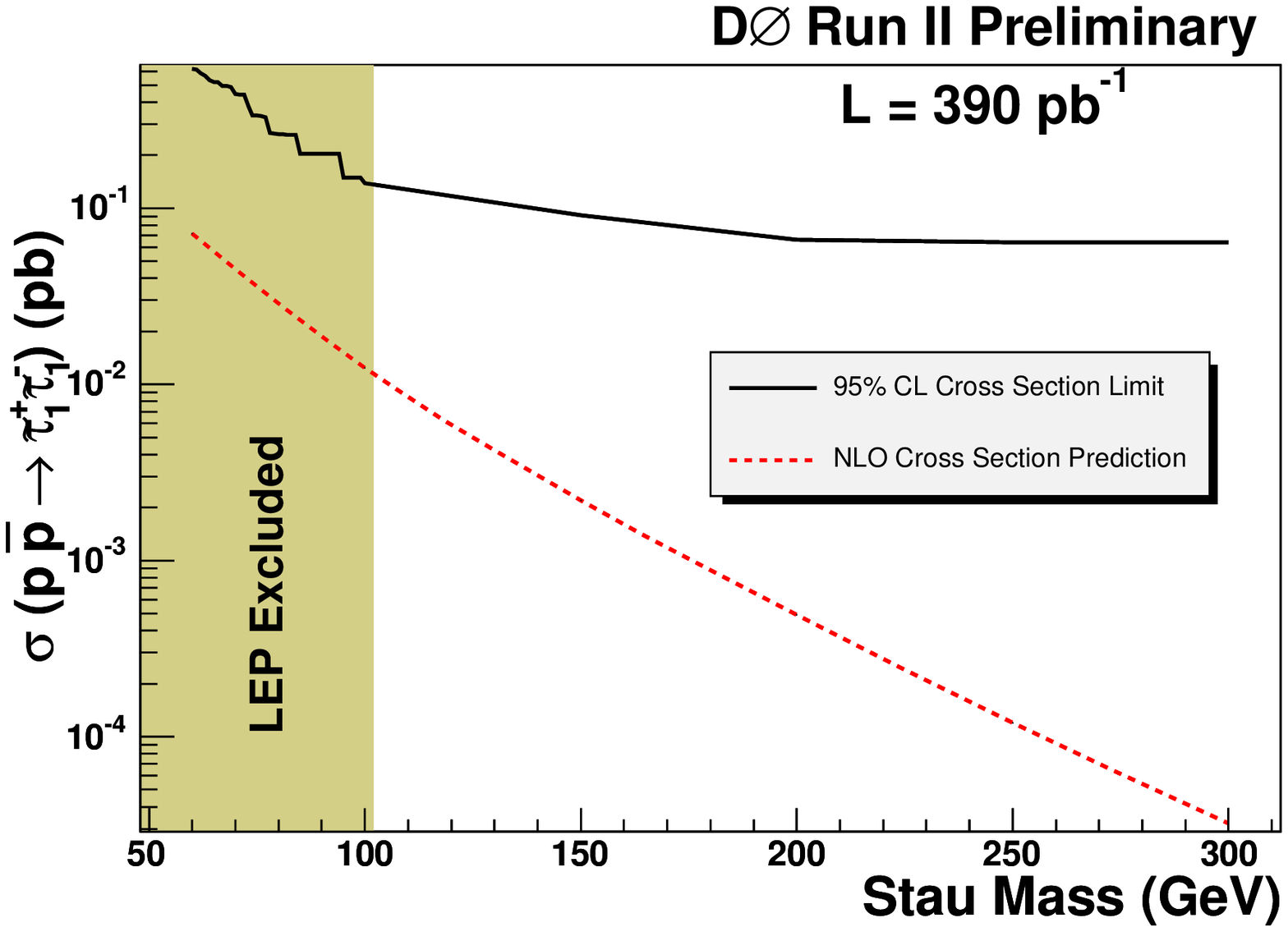}
\includegraphics[width=.5\textwidth]{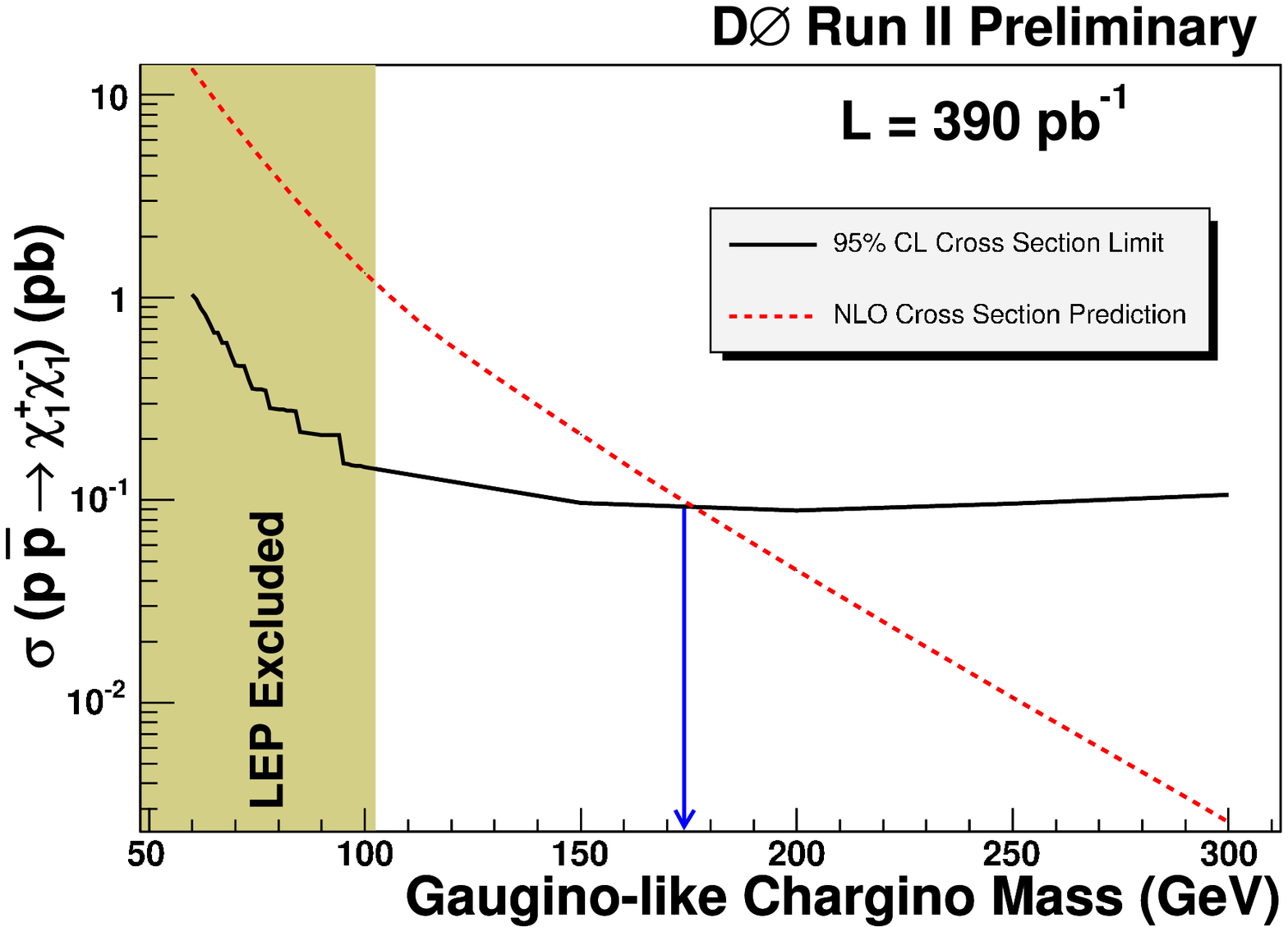}
\caption{95\% cross-section limit (solid line) and NLO production cross-section
for pair-produced staus in GMSB (left) and gaugino-like charginos (right).}
\label{fig2}
\end{figure}

95\% confidence level limits on the cross-section for CMSP pair-production 
are set
using the $CL_s$ method~\cite{Junk} for each model and CMSP mass point 
and are subsequently compared to
the next-to-leading order cross-section predictions calculated with 
Prospino~2~\cite{Beenakker}.
In the case of the GMSB analysis, these limits are not yet stringent enough to set
a limit on the stau mass, nevertheless they are the best limits to date from
the Tevatron (cf. Fig.\ref{fig2}, left).

The cross-section limits for the chargino-analyses are shown in 
Fig.\ref{fig2}, right.
Mass limits for the stable chargino could be set beyond the LEP exclusion 
limit of 102.5\,GeV~\cite{lepsusy}:
For higgsino-like stable charginos masses below 140\,GeV, for
gaugino-like stable charginos masses below 174\,GeV are excluded.
These are currently the best mass limits
on stable charginos.

\end{document}